\documentclass[journal]{IEEEtran}

\usepackage{graphics} 
\usepackage{graphicx}
	\graphicspath{{figure/}}
	\DeclareGraphicsExtensions{.jpeg,.png}
\usepackage{array}
\usepackage{verbatim}
\usepackage{amsmath}
\usepackage{amssymb}
\usepackage{caption}
\usepackage{subcaption}
\usepackage{color,soul}

\usepackage{multirow}
\usepackage{booktabs}

\usepackage{cite}

\makeatletter
\def\hlinewd#1{%
\noalign{\ifnum0=`}\fi\hrule \@height #1 %
\futurelet\reserved@a\@xhline}
\makeatother

\title{
Online Mental Stress Detection using Frontal-channel EEG Recordings in a Classroom Scenario
}

\author{
	Chi-Yuan Chang,
	Chieh Hsu,
        Ying Choon Wu,
        Siwen Wang,
        Darin Tsui
	and Tzyy-Ping Jung,~\IEEEmembership{Fellow,~IEEE}

\thanks{This work was supported in part by a research contract from Facebook, a gift fund, KreutzKamp TMS RES F- 2467, to TPJ,  the Army Research Laboratory (W911NF-10-2-0022) and the National Science Foundation (DUE-1734883).}
	\thanks{C.-Y. Chang is with Harvard Medical School and Beth Israel Deaconess Medical Center, Boston (BIDMC).}
     \thanks{Y.-C. Wu and T.-P. Jung are with Dept. of Bioengineering (BIOE) and Swartz Center for Computational Neuroscience (SCCN) of the University of California, San Diego (UCSD).}
}

\begin{document}
\bstctlcite{IEEEexample:BSTcontrol}
\maketitle
\thispagestyle{empty}
\pagestyle{empty}
\begin{abstract}

\textit{Objective:} To investigate the effects of different approaches to EEG preprocessing, channel montage selection, and model architecture on the performance of an online-capable stress detection algorithm in a classroom scenario.
\textit{Methods:} This analysis used EEG data from a longitudinal stress and fatigue study conducted among university students. The students' self-reported stress ratings during each class session were the basis for classifying EEG recordings into either normal or elevated stress states.
We used a data-processing pipeline that combined Artifact Subspace Reconstruction (ASR)and an Independent Component Analysis (ICA)-based method to achieve online artifact removal.
We compared the performance of a Linear Discriminant Analysis (LDA) and a 4-layer neural network as classifiers.
We opted for accuracy, balanced accuracy, and F1 score as the metrics for assessing performance.
We examined the impact of varying numbers of input channels using different channel montages. Additionally, we explored different window lengths and step sizes during online evaluation.
\textit{Results:} 
Our online artifact removal method achieved performance comparable to the offline ICA method in both offline and online evaluations.
A balanced accuracy of 77\% and 78\% in an imbalanced binary classification were observed when using the 11-frontal-channel LDA model with the proposed artifact removal method.
Moreover, the model performance remained intact when changing the channel montage from 30 full-scalp channels to just 11 frontal channels.
During the online evaluation, we achieved the highest balanced accuracy (78\%) with a window length of 20 seconds and a step size of 1 second.
\textit{Conclusions:} In this study, we present an EEG-based algorithm for online mental stress detection. It incorporates an artifact removal pipeline that combines ASR with an ICA.
The algorithm employs an 11-frontal-channel LDA model, showcasing promising potential for real-world stress monitoring applications.
\textit{Significance:} 
This study comprehensively investigates the deployment of stress detection in real-world scenarios.
The findings of this study provide insight into the development of daily mental stress monitoring.

\end{abstract}
\section{Introduction}
\IEEEPARstart{S}{\lowercase{tress}} is a reaction elicited when the demands from the environment exceed the coping capacity of an organism, resulting in physiological and biological adaptive responses related to fight or flight. \cite{cohen1997stress_def}.
These demands are called stressors and can be further categorized into physical/ environmental or mental/ task-related stressors \cite{giannakakis2019stress_review_physiological_change}.
In our daily lives, we face various challenges and experience different levels of mental stress.
However, people usually underestimate the effect of mental stress on well-being, both physically and mentally.
When experiencing low-level stress, people might suffer from upset stomachs \cite{behere2011stress_upset_stomach}, dizziness \cite{andersson2000stess_dizzy}, increased heart rate \cite{delaney2000stress_hr}, or muscle tension \cite{lundberg1999stress_muscle}.
As stress levels increase, the symptoms can become more severe.
Further,  researchers have linked mental stress to various health issues, such as immune disorders \cite{segerstrom2004immune}, obesity \cite{jarali2013obesity}, and heart disease \cite{pickering2001mental_stress_heart_disease}.
Additionally, individuals subject to chronic stress are prone to neuropsychiatric conditions, including depression \cite{checkley1996stress_depression} and anxiety \cite{dieleman2015stress_anxiety}, as well as stroke \cite{kotlkega2016stress_stroke}.
 In extreme cases, mental stress can lead to substance abuse and suicide \cite{brent1995stress_substance_abuse}.
Therefore, mental stress has been recognized as the top proxy killer disease \cite{debnath2011proxy}.

While recognizing its negative impacts, we do not necessarily need to avoid mental stress \cite{selye1976stress_is_not_bad}.
In 1908, Yerkes and Dodson proposed a model to explain the relationship between arousal and task performance \cite{yerkes1908stress_principle, staal2004stress_curve}.
 The model indicates that performance improves as arousal increases in simple task conditions, such as focused attention, flashbulb memory, and fear conditioning. Conversely, in challenging task conditions, such as divided attention, working memory, and decision-making, the relationship between arousal and task performance follows an inverted U shape.
Because stress levels rise concurrently with arousal, we can model the relationship between stress and task performance similarly. Indeed, research has shown that moderate levels of mental stress can enhance performance in competitive sports\cite{jones1989stress_sport} and certain attention-related tasks \cite{staal2004stress_curve}.


Many of the tasks we encounter in daily life are challenging and stressful, making monitoring mental stress essential for maximizing performance. Questionnaires and self-reports have long served as tools for measuring stress levels \cite{cahir1991PSSQ_stress_questionnaire, crandall1992USQ_stress_questionnaire, levenstein1993perceived_stress_questionnaire, lovibond1995_dass21}, considering them as the standard for stress level determination in related studies \cite{rissen2000stress_EMG_selfreport, lopez2012stress_frontal_asymmetry_selfreport, blackhart2006stress_frontal_asymmetry_selfreport}.
However, real-time monitoring of mental stress using questionnaires and self-reports is not feasible. Furthermore, these measurements are subjective and may be susceptible to anchoring effects  \cite{paulhus2007self-report-guide}.

In contrast, physiological signals provide objective measures for stress detection. 
The Automatic Nervous System (ANS) controls the involuntary movements of the body, such as skin conductivity, heart rate, and pupil dilation \cite{langley1921ANS, sharma2012stress_measure, beatty2000stress_pupil}.
Since stress causes dynamic changes in the ANS, there are several techniques for detecting stress level changes \cite{giannakakis2019stress_review_physiological_change, sharma2012stress_measure}.
The most commonly used signals in stress detection are Heart Rate (HR), Heart Rate Variability (HRV), Galvanic Skin Response (GSR), blood pressure, and respiration rate. \cite{Shruti2021_review_stress_detection}.
Previous studies have shown that, under stress, HR, GSR, blood pressure, and respiration rate increase, and the R-peak-to-R-peak interval in HRV shorten \cite{giannakakis2019stress_review_physiological_change, sharma2012stress_measure}.
Even though HR has been used extensively in previous studies, HR might not accurately reflect mental stress in situations involving body movement, as HR increases during exercise.

Electroencephalography (EEG), a non-invasive measure of cortical activity, is another commonly used method for stress detection \cite{sharma2012stress_measure}.
Since EEG offers a high temporal resolution and portability and is lower in cost than other neuroimaging techniques, it has been widely used in neuroscience \cite{grummett2014_gamma,Artoni2017}, clinical assessment \cite{marzbani2016_review_BCN}, and brain-computer interfaces \cite{Fabiani2004}.
\begin{table*}[ht!]
\centering
\caption[Previous studies of stress detection.]{Previous studies of stress detection.}
\begin{tabular*}{\textwidth}{@{\extracolsep{\fill}}l|c c c}
    \specialrule{1.5pt}{0pt}{0pt}
    \textbf{Study}
    & \textbf{Stimuli}
    & \textbf{Biosignals used}
    & \textbf{Accuracy}
    \\ \hline
    Xia et al.\cite{xia2018stress_detection_mental_arithmetic_task}
    & Mental Arithmetic Task
    & EEG, ECG
    & 79.54\%
    \\ \hline
    Minguillon et al.\cite{minguillon2018stress_detection_montreal_imaging}
    & Montreal Imaging Stress Task
    & EEG, ECG, EMG, EDA
    & 94.60\%
    \\ \hline
    Asif et al.\cite{asif2019stress_detection_music_tracks}
    & Music Tracks
    & EEG
    & 98.76\%
    \\ \hline
    Saeed et al.\cite{saeed2020stress_detection_long-term}
    & No stimuli
    & EEG
    & 85.20\%
    \\
    \specialrule{1.5pt}{0pt}{0pt}
\end{tabular*}
\label{previous_studies}
\end{table*}

Compared to commonly used ANS-based measures of stress, EEG provides a more comprehensive,  detailed, and temporally sensitive picture of stressor impacts over time by measuring changes in the oscillatory activity of the cortex.
Hence, EEG can open new possibilities for stepping outside of the laboratory and tackling the challenge of real-world stress as well. However, although many studies have shown the feasibility of detecting stress in laboratory environments using EEG and other measures, stress detection in realistic and real-world contexts is still in its infancy. As shown in Table \ref{previous_studies}, stress is usually studied through artificial tasks that do not necessarily resemble experiences of stressors encountered in day-to-day life \cite{saeed2020stress_detection_long-term}.

To approach this question, this study analyzed EEG data from university students during regular coursework, including lectures, quizzes, and other activities, for a semester   \cite{ko2017classroom_data}. Students' ratings of their subjective stress during each class session served as ground truth for classifying each session of EEG recordings as normal or elevated stress. We explore different approaches to EEG preprocessing, channel selection, and model architecture in service of an online-capable stress detection algorithm.


\section{Materials and Methods}
\subsection{Data recording and preprocessing}
\subsubsection{Data recording}
At the beginning of each lecture, students were asked to complete the Depression, Anxiety, Stress Scales-21 (DASS-21) questionnaire \cite{lovibond1995_dass21}.
Next, eye-open baseline EEG and ECG were recorded for about five minutes using a 32-channel NeuroScan System at a 1000 Hz sampling rate. Subsequently, the students normally participated in their lectures and other classroom activities for about sixty minutes while their EEG and ECG were continuously recorded. After class, we recorded 5 minutes of resting state eye-open EEG and ECG a second time.
\par
Eighteen subjects participated in the first semester, and eight new subjects in the second semester.
We recorded 4 to 13 sessions per subject on different days.
Because of insufficient sessions from the second semester, we analyzed only the data from the first semester.
\subsubsection{Data preprocessing}
To classify mental stress levels using non-task-related data, we first extract the 5-minute eye-open resting EEG at the beginning of the experiment.
Then, we performed band-pass filtering from 1Hz to 50Hz to remove high-frequency noise.
We further removed bad channels that contained negligible activity, noisy signals, or have a poor correlation with adjacent channels.
Next, we implemented Artifact Subspace Reconstruction (ASR) to reconstruct the portion where EEG signals are contaminated by muscle-related or motion-related artifacts \cite{Kothe2014, chang2019evaluation}.
Finally, we implemented Independent Component Analysis (ICA) and rejected ICs with a probability higher than 80\% of being artifact-related ICs labeled by \texttt{ICLabel} \cite{pion2019_iclabel}.

\subsection{Data labeling and feature extraction}
\subsubsection{Data labeling}
We used the stress index in the DASS-21 questionnaire to quantify students' subjective stress, which ranged from 0 to 38 across all sessions. 
DASS-21 scores divided each session into elevated or normal stress groups, with scores above 14 representing elevated stress levels.
After removing the sessions missing DASS-21 reports, data remained for 71 normal and 21 elevated stress sessions.


\subsubsection{Feature extraction}
Based on 50 frequency bins with a bin width of 1Hz, we estimated the power spectrum density (PSD) in the eye-open resting sessions.
To reduce the number of features, we selected frequency bins in the $\theta$ range as $\theta$ band activities have been linked to stress-related cognitive state changes in previous studies \cite{sani2014stress_detection_power, alonso2015stress_theta_increase}.
We did not analyze activities in frequency ranges because they overlapped with other non-brain artifacts, such as head movement \cite{goncharova2003emg_freq_range}, or did not reflect consistent stress-induced brain response.
The final number of features used in the analysis was 150 (5 frequency bins $\times$ 30 channels).

\subsubsection{Synthetic minority over-sampling technique}
Training classifiers on the imbalanced classes can potentially bias results toward the majority group.
To alleviate the imbalanced classes problem, we evaluated the performance of the synthetic minority over-sampling technique (SMOTE) \cite{chawla2002smote}, which essentially interpolates data points in the smaller group to create balanced class sizes.
It does so by first randomly selecting one data point from the minority group.
Next, it calculates k-nearest neighbors in the minority group of the selected data point.
Finally, SMOTE randomly selects one neighbor and generates a synthetic minority data point between the selected data point and the neighbor.
By repeating this process, the number of data points in the majority and minority groups is equal.

\subsection{Model selection and evaluation}
\subsubsection{Linear Discriminant Analysis}
\label{lda_method}
The first classification model we investigated is Linear Discriminant Analysis (LDA) \cite{balakrishnama1998LDA}, 
which aims to find a linear combination of features to separate different classes into different distributions.
In this study, we implement LDA in \texttt{scikit-learn} \cite{scikit-learn}, a Python library, with equal prior probabilities for both classes.
Since LDA is sensitive to outliers \cite{buyukozturk2008lda_outlier}, we further selected 3Hz to 7Hz from Fz, FCz, and Cz channels out of 150 features to prevent potential outlier effects.
These channels lie along the mid-line of the scalp topology, and previous studies have shown that they typically capture activities closely associated with changes in cognitive states within the \(\theta\) range \cite{attallah2020frontal_electrode}.

\subsubsection{Neural Network}
\begin{figure}
    \centering
    \includegraphics[width=1\linewidth]{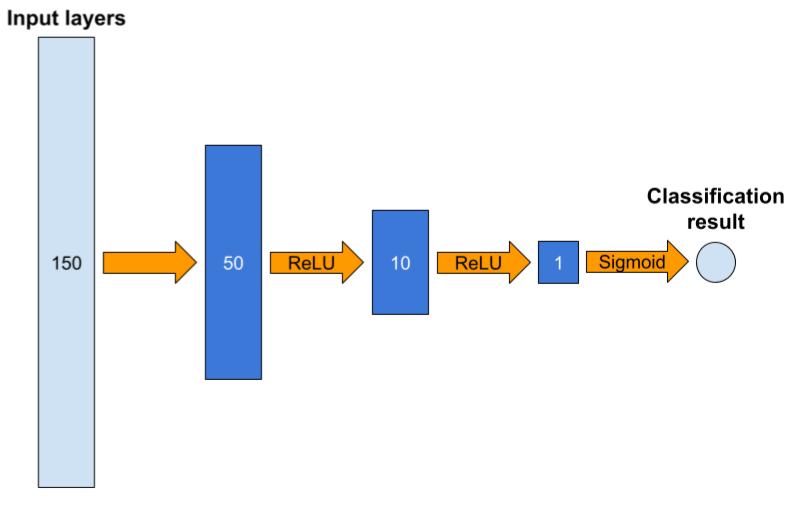}
    \caption[The architecture of the Neural Network]{The architectures of the Neural Network used in this study. The number of nodes is indicated by the number in each layer. All the layers are fully connected to the next layer.}
    \label{nn_struct}
\end{figure}

Besides the linear classifier, we also investigated the performance of A non-linear classifier.
In particular, we built a 4-layer fully connected neural network (NN) using \texttt{pyTorch} \cite{NEURIPS2019_9015pytorch} and trained this NN using the features from all the recording channels.
Fig.~\ref{nn_struct} depicts the architectures of the NN.
We used ReLU as an activation function between layers and a Sigmoid function at the output layers.
The number of training epochs is 1000, and the loss function for backpropagation is binary cross-entropy loss.

\subsubsection{Model evaluation}
To evaluate the performance of our model, we implemented leave-one-session-out (LOO) cross-validation after model training.
LOO cross-validation selects a test session from the dataset and then trains the model on the rest of the sessions. Next, the trained model is applied to the test session. LOO cross-validation performs the same procedure for all sessions and finally reports the prediction results.
\par
Since our dataset contains more normal than elevated stress sessions, the chance level calculated over all the sessions is 77.17\%.
Therefore, instead of using accuracy as a performance metric, we used balanced accuracy, which gives us a chance level of 0.5.
We also used the F1 Score to indicate the harmonic balance between recall and precision to characterize a model's sensitivity and specificity.



\subsection{Practical issues of deployment}
We must address several practical issues before deploying stress-detection systems in the real world.
First, we would like to monitor our stress levels in real time.
Unlike the laboratory scenarios, asking the users to wait 5 minutes before getting the results is unrealistic.
Hence, online capability is one of the most important indicators when we build our stress detection.
Second, we would like to reduce the number of recording channels needed in our stress detection because EEG setup can be time-consuming and labor-intensive.
Adding additional channels to a recording system may also increase its manufacturing cost significantly.
Therefore, we investigated the effect of recording channel reduction on our stress detection algorithms.

\subsubsection{Real-time ocular artifact removal}
Independent component analysis can identify artifact-related ICs, especially eye-related ICs, and remove them from the EEG signals \cite{Jung2000_bbs, bigdely2013eyecatch, Frolich2004}.
However, the computational time of ICA makes it unsuitable for real-time applications.
This study proposes an IC projection artifact removal method for removing eye-related artifacts without performing online ICA.
We focus solely on removing eye-related ICs for two key reasons. Firstly, eye activities introduce contamination into EEG signals within the frequency range of 0Hz to 8Hz, which coincides with the frequency band of our PSD features \cite{schlogl2007eog_freq_range}.
Second, compared to  sparse and complicated muscle activities, two kinds of the most severe eye artifacts, blink and horizontal saccade, can be well separated by ICA into blink- and saccade-related ICs.  Furthermore, these eye-related ICs exhibit remarkable similarity across subjects.
We performed ICA for each session separately to check whether most contained blink- and saccade-related ICs. We counted the number of eye-related ICs found in each session.
\par
Our proposed method is comprised of three steps.
First, we performed ICA on a single subject and recorded its ICA unmixing matrix $W$ and its ICA mixing matrix $W^{-1}$.
Next, we removed the blink- and saccade-related ICs from the mixing matrix $W^{-1}$ by assigning the corresponding columns to 0 and recorded this new mixing matrix $M^*$.
In cases where there is a discrepancy in the number of recording channels between the template and the processed data, we approximated the new mixing and unmixing matrices by retaining the rows in $M^*$ and the columns in $W$ according to the channel labels.
Finally, we combined the unmixing matrix and the new mixing matrix to create an IC decomposition template $P$.
This template can remove the eye-related IC activities using matrix multiplication, thereby preventing the need for online ICA execution.


\subsubsection{Reduction of recording channels}
\label{sec_reduce_ch}
\begin{figure}
    \centering
    \includegraphics[width=1\linewidth]{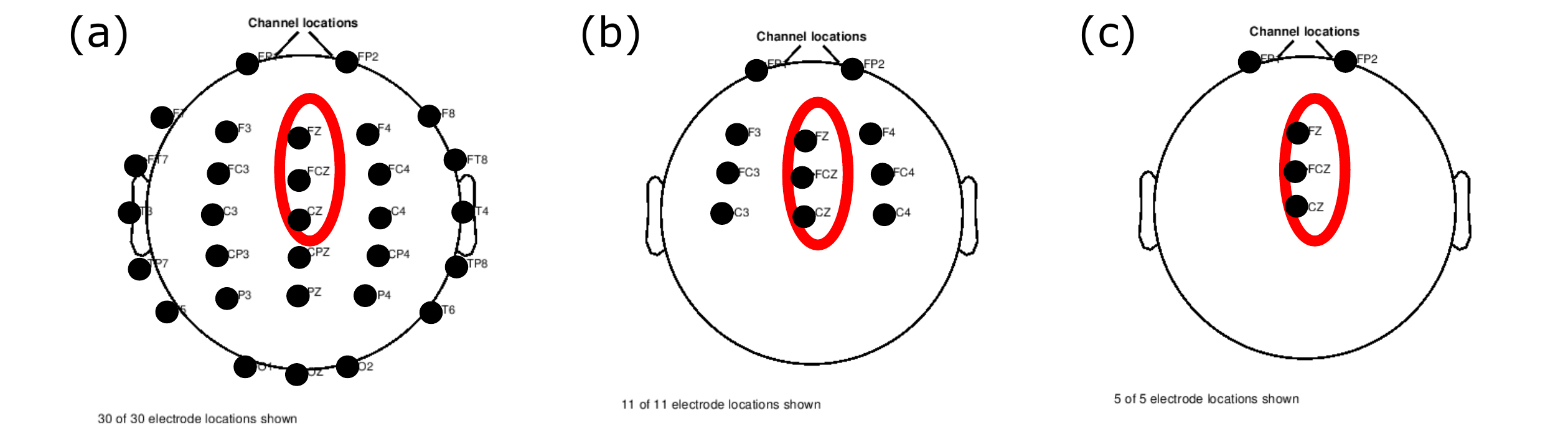}
    \caption[Montage of channel reduction.]{Montage of channel reduction. (a) shows all 30 channels. (b) shows the subselected 11 channels. (c) shows the subselected 5 channels. The red circle indicates the location of 3 feature channels.}
    \label{stress_ch}
\end{figure}

We compared three recording montages and their stress-detection performance to investigate the effect of recording channel reduction.
Besides the original 30 recording channels, we selected 11 and 5 channels in the frontal region out of the 30 channels.
Fig.~\ref{stress_ch} shows the location of the three channels.
We selected these channels because we would like to capture the eye-related ICs in the IC decomposition template.
After the channel reduction, we performed data preprocessing with the IC projection artifact removal method and trained our stress detection algorithms with LDA and NN.
One thing to clarify is that the number of nodes in the input layer of NN changes from 150 to 55 and 25, corresponding to 11 channels and 5 channels, to adapt to the change in the number of recording channels.

\subsubsection{Online stress detection evaluation}
\label{sec_online_stress}
We investigated two important parameters in online use cases: the length of the sliding window for the PSD estimation and the step size to move the sliding window forward.
First, we segmented a session into multiple epochs using different sliding window lengths and step sizes.
Next, we trained our LDA stress detection model using the rest of the sessions and applied the model to the epochs.
For each window, we had a prediction about whether the session was in the elevated stress group or the normal stress group.
To summarize these predictions, we used a majority vote to assign the session to the group with the most predictions.
After repeating this process for all the sessions, we calculated the balanced accuracy for each pair of sliding window lengths and step sizes.

\section{Results}
\subsection{Stress detection with and without artifact removal and SMOTE}

\begin{figure}
 	\centering
 	\includegraphics[width=1\linewidth]{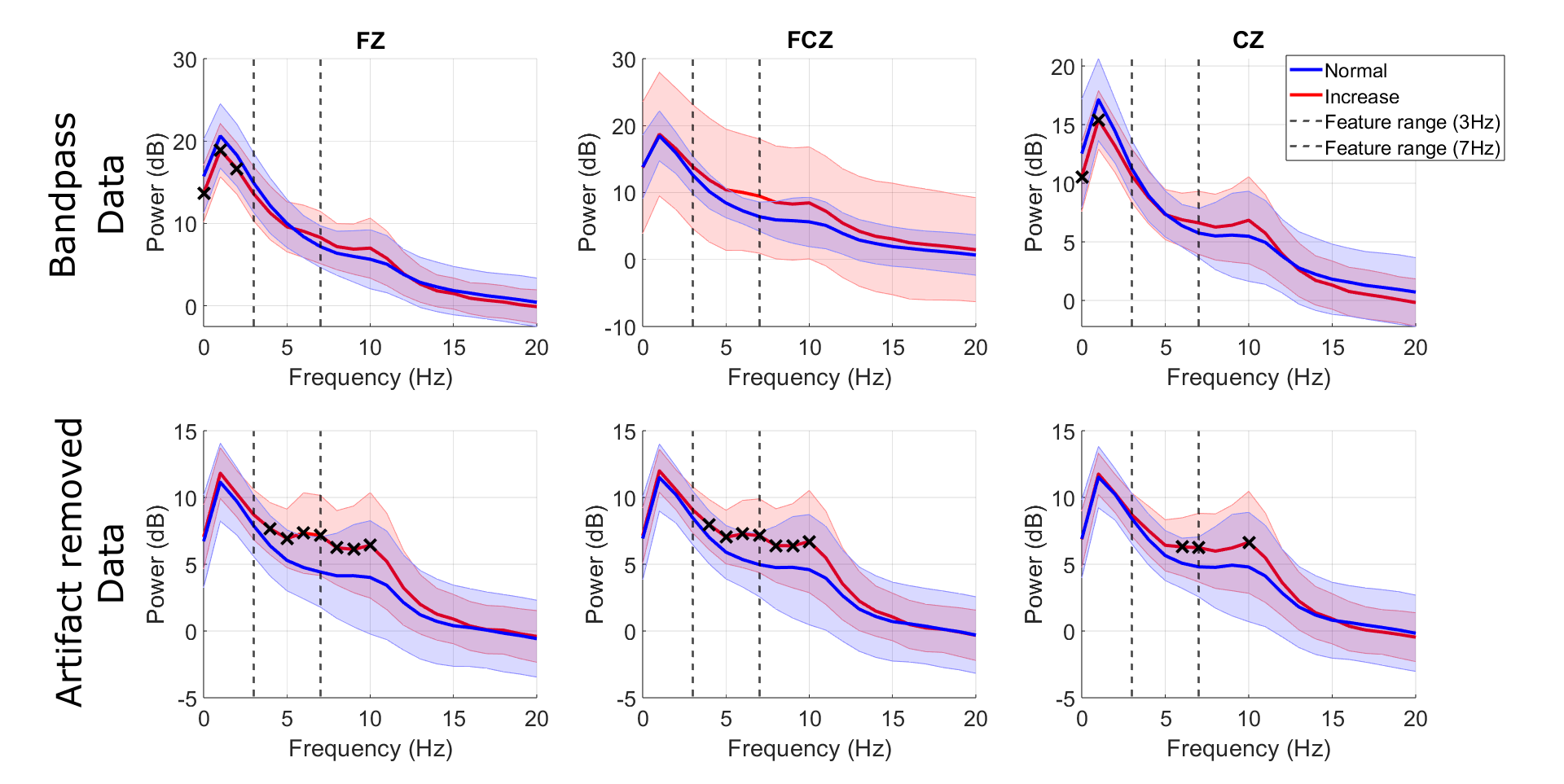}
 	\caption[PSD features of selected channels before and after artifact removal]{PSD features of selected channels before and after artifact removal. The first row shows the PSDs of band-pass-filtered only data. The second row shows the PSDs of data that have undergone artifact removal. The spectral activities of the normal and elevated stress groups are shown in blue and red, respectively. The solid lines indicate the mean, and the shaded area indicates the standard deviation across sessions. The cross dots indicate frequency bins in which the spectral power of EEG activities reliably differentiates the two groups using a bootstrap test ($p<0.05$). The dashed line indicates the frequency range used in the LDA analysis.}
 	\label{cmp_psd}
\end{figure}

Fig.~\ref{cmp_psd} shows the effect of our artifact-removal pipeline on the PSD features of selected channels.
On the top row, the PSDs of the elevated and normal stress groups  T after band-pass filtering alone do not directly differentiate in the feature frequency range of all the feature channels.
On the other hand, after artifact removal, we found noticeable group differences between 4 to 10Hz over all three channels (bottom row). The scale of the PSD also dropped from 20 dB to 10dB after artifact removal.

The comparisons of stress detection performance before and after artifact removal are shown in Table~\ref{stress_lda_artifact} and Table~\ref{stress_nn_artifact} concerning the LDA model and NN model.
In both models, we observe a gradual increase in accuracy and balanced accuracy after each artifact removal step.
Also, the performance of the LDA model is always better than the NN model.
The best performance appears when combining ASR with offline ICA.
When replacing the offline ICA method with the IC projection artifact removal method, we observe noticeable improvements compared to applying ASR alone in both models.
Moreover, there is only a 1\% difference between the LOO balanced accuracy of the proposed method and offline ICA in the LDA model.

\begin{table}[b]
\centering
\caption[The effects of artifact removal on LDA stress detection]{The effects of artifact removal on the performance of LDA-based stress detection. BP refers to band-pass filtering. Projection refers to the IC projection artifact removal method.}
\begin{tabular*}{\linewidth}{@{\extracolsep{\fill}} l|c c c}
    \specialrule{1.5pt}{0pt}{0pt}
    \multicolumn{1}{c|}{\textbf{LDA}}
    & \textbf{LOO Acc.}
    & \textbf{LOO Balanced Acc.}
    & \textbf{F1}
    \\
    \specialrule{1.5pt}{0pt}{0pt}
    BP Only
    & 68.48\%
    & 62.81\%
    & 0.43
    \\ \hline
    BP + ASR
    & 72.83\%
    & 70.66\%
    & 0.53
    \\ \hline
    \textbf{BP + ASR + ICA}
    & \textbf{79.35\%}
    & \textbf{78.24\%}
    & \textbf{0.63}
    \\ \hline
    BP + ASR + Projection
    & 80.43\%
    & 77.26\%
    & 0.63
    \\ \specialrule{1.5pt}{0pt}{0pt}
\end{tabular*}
\label{stress_lda_artifact}
\end{table}

\begin{table}[t]
\centering
\caption[The effects of artifact removal on 4-layer NN stress detection]{The effects of artifact removal on the performance of 4-layer NN stress detection. BP refers to band-pass filtering. Projection refers to the IC projection artifact removal method.}
\begin{tabular*}{\linewidth}{@{\extracolsep{\fill}} l|c c c}
    \specialrule{1.5pt}{0pt}{0pt}
    \multicolumn{1}{c|}{NN}
    & LOO Acc.
    & LOO Balanced Acc.
    & F1
    \\
    \specialrule{1.5pt}{0pt}{0pt}
    BP Only
    & 75.00\%
    & 56.98\%
    & 0.3
    \\ \hline
    BP + ASR
    & 77.17\%
    & 66.77\%
    & 0.49
    \\ \hline
    \textbf{BP + ASR + ICA}
    & \textbf{83.70\%}
    & \textbf{74.35\%}
    & \textbf{0.62}
    \\ \hline
    BP + ASR + Projection
    & 80.43\%
    & 68.88\%
    & 0.53
    \\
    \specialrule{1.5pt}{0pt}{0pt}
\end{tabular*}
\label{stress_nn_artifact}
\end{table}

Since our two groups of data did not contain balanced numbers of datasets, we evaluated whether SMOTE could enhance the performance of our stress detection models.
Table~\ref{stress_smote} shows the effect of SMOTE on the analysis after the artifact removal pipeline with offline ICA.
Though the LOO balanced accuracy increased slightly, there was no significant improvement in both models with the SMOTE process.
\par
In the following analysis, we only reported the results from the LDA model without SMOTE, as the LDA model consistently outperformed the NN model, and the effect of SMOTE was found to be negligible.

\begin{table}[t]
\centering
\caption[The effects of SMOTE on LDA and NN stress detection]{The effect\st{s} of SMOTE on LDA and NN stress detection.}
\begin{tabular*}{\linewidth}{@{\extracolsep{\fill}} l|c c c}
    \specialrule{1.5pt}{0pt}{0pt}
    & LOO Acc.
    & LOO Balanced Acc.
    & F1
    \\
    \specialrule{1.5pt}{0pt}{0pt}
    LDA (w/o SMOTE)
    & 79.35\%
    & 78.24\%
    & 0.63
    \\ \hline
    LDA (w/ SMOTE)
    & 81.52\%
    & 79.64\%
    & 0.65
    \\ \hline
    NN (w/o SMOTE)
    & 83.70\%
    & 74.35\%
    & 0.62
    \\ \hline
    NN (w/ SMOTE)
    & 81.52\%
    & 76.29\%
    & 0.62
    \\
    \specialrule{1.5pt}{0pt}{0pt}
\end{tabular*}
\label{stress_smote}
\end{table}

\subsection{Consistency of eye-related ICs across different recordings}
\begin{figure}
    \centering
    \includegraphics[width=0.7\linewidth]{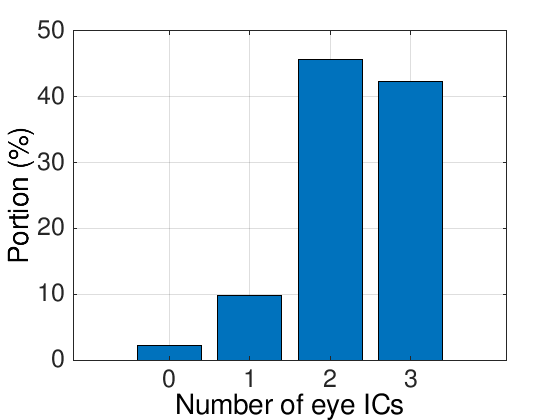}
    \caption[Histogram of the number of eye-related ICs found in each session.]{Histogram of the number of eye-related ICs found in each session. The histogram has been normalized by the number of sessions.}
    \label{nb_eyeIC_portion}
\end{figure}

\begin{figure}
     \centering
     \includegraphics[width=0.8\linewidth]{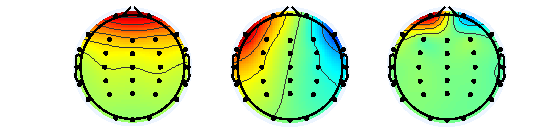}
     \caption[Example of the three eye-related ICs found within a session.]{Example of the three eye-related ICs found within a session. The left and center ICs correspond to blink- and saccade-related ICs, respectively.}
     \label{ex_3_eye}
 \end{figure}

To evaluate the consistency of eye-related ICs across different recordings, Figure.~\ref{nb_eyeIC_portion} shows the portion of sessions with different numbers of eye-related ICs found.
We found that 45.7\% of sessions contain two eye-related ICs, and 42.4\% of sessions contain three eye-related ICs.
Moreover, only 2.2\% of sessions contain no eye-related ICs.
Fig.~\ref{ex_3_eye} shows an example of the three eye-related ICs in a session.

 \begin{figure}
    \centering
    \includegraphics[width=1\linewidth]{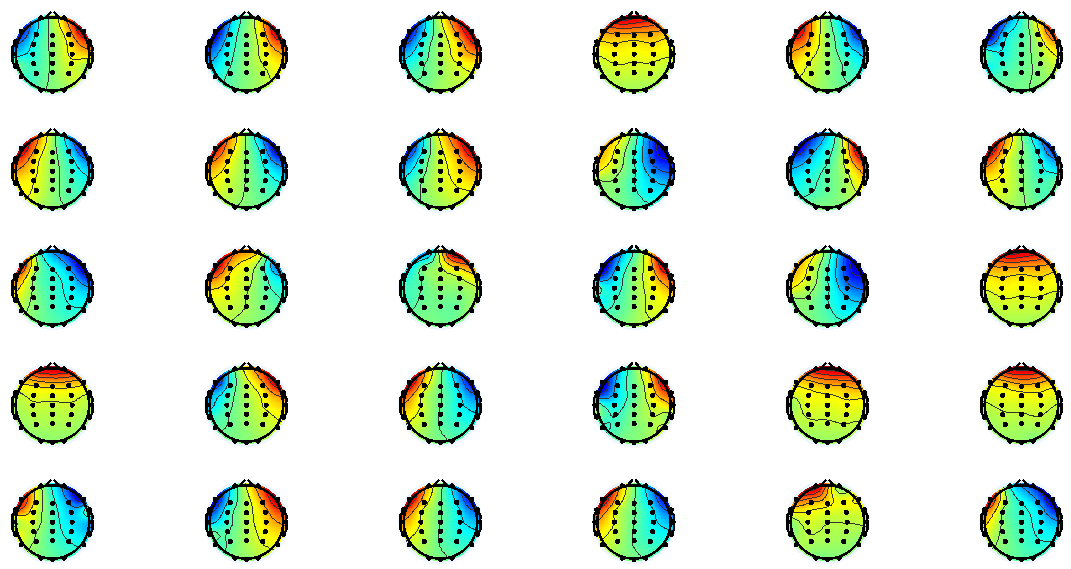}
    \caption[Example of the rejected eye-related ICs.]{Example of the rejected eye-related ICs.}
    \label{ex_rej_eye}
\end{figure}

\subsection{Effect of reducing the channel montage on stress level classification accuracy}
After evaluating the performance of the IC projection artifact removal method, we further compared the effect of reducing the number of recording channels.
Table~\ref{stress_lda_ch} shows the performance of the LDA models trained on data recorded by different input channels.
All data were processed using the IC projection artifact removal method before training LDA models.
We found that the performance of the LDA classifier increased by 1\% after the electrode montage was reduced from 30 to 11 channels.
However, with a 5-channel montage, the balanced accuracy dropped drastically from 77\% to 60\%.

\begin{table}[t]
\centering
\caption[The effects of channel reduction on LDA-based stress detection]{The effects of reducing input channels on the performance of LDA-based stress detection. After reducing the channel montage, the data are preprocessed using the proposed projection method. All the models used frequency bins between 3 to 7Hz on Fz, FCz, and Cz as features.}
\begin{tabular*}{\linewidth}{@{\extracolsep{\fill}} l|c c c}
    \specialrule{1.5pt}{0pt}{0pt}
    \multicolumn{1}{c|}{LDA}
    & LOO Acc.
    & LOO Balanced Acc.
    & F1
    \\
    \specialrule{1.5pt}{0pt}{0pt}
    30 Channels
    & 80.43\%
    & 77.26\%
    & 0.63
    \\ \hline
    \textbf{11 Channels}
    & \textbf{80.43\%}
    & \textbf{78.94\%}
    & \textbf{0.64}
    \\ \hline
    5 Channels
    & 67.39\%
    & 60.43\%
    & 0.40
    \\
    \specialrule{1.5pt}{0pt}{0pt}
\end{tabular*}
\label{stress_lda_ch}
\end{table}

\begin{table*}[t]
\centering
\caption[Leave-one-session-out validation of LDA stress detection]{Leave-one-session-out validation of LDA stress detection. All the models used frequency bins 3Hz to 7Hz on Fz, FCz, Cz as features. \textit{Proj}. refers to the IC projection artifact removal method.}
\begin{tabular*}{\textwidth}{@{\extracolsep{\fill}} l|l c c c}
    \specialrule{1.5pt}{0pt}{0pt}
    \multicolumn{1}{c|}{LDA}
    & Preprocessing
    & LOO Acc.
    & LOO Balanced Acc.
    & F1
    \\ \specialrule{1.5pt}{0pt}{0pt}
    & BP Only
    & 68.48\%
    & 62.81\%
    & 0.43
    \\ \hline\hline
    \multirow{3}{*}{30 Channels} & \multicolumn{1}{l}{BP+ASR} & \multicolumn{1}{c}{72.83\%} & \multicolumn{1}{c}{70.66\%} & \multicolumn{1}{c}{0.53} \\\cline{2-5}
    & \multicolumn{1}{l}{BP+ASR+ICA} & \multicolumn{1}{c}{79.35\%} & \multicolumn{1}{c}{78.24\%} & \multicolumn{1}{c}{0.63} \\\cline{2-5}
    & \multicolumn{1}{l}{BP+ASR+Proj.} & \multicolumn{1}{c}{80.43\%} & \multicolumn{1}{c}{77.26\%} & \multicolumn{1}{c}{0.63} \\\hline\hline
    \multirow{2}{*}{11 Channels} & \multicolumn{1}{l}{BP+ASR} & \multicolumn{1}{c}{75.00\%} & \multicolumn{1}{c}{75.42\%} & \multicolumn{1}{c}{0.58} \\\cline{2-5}
    & \multicolumn{1}{l}{\textbf{BP+ASR+Proj.}} & \multicolumn{1}{c}{\textbf{80.43\%}} & \multicolumn{1}{c}{\textbf{78.94\%}} & \multicolumn{1}{c}{\textbf{0.64}} \\\hline\hline
    \multirow{2}{*}{5 Channels} & \multicolumn{1}{l}{BP+ASR} & \multicolumn{1}{c}{70.65\%} & \multicolumn{1}{c}{64.22\%} & \multicolumn{1}{c}{0.45} \\\cline{2-5}
    & \multicolumn{1}{l}{BP+ASR+Proj.} & \multicolumn{1}{c}{67.39\%} & \multicolumn{1}{c}{60.43\%} & \multicolumn{1}{c}{0.40} 
    \\ \specialrule{1.5pt}{0pt}{0pt}
\end{tabular*}
\label{lo-session-o summary}
\end{table*}

\subsection{Online stress detection}
\begin{figure*}
    \centering
    \includegraphics[width=1\linewidth]{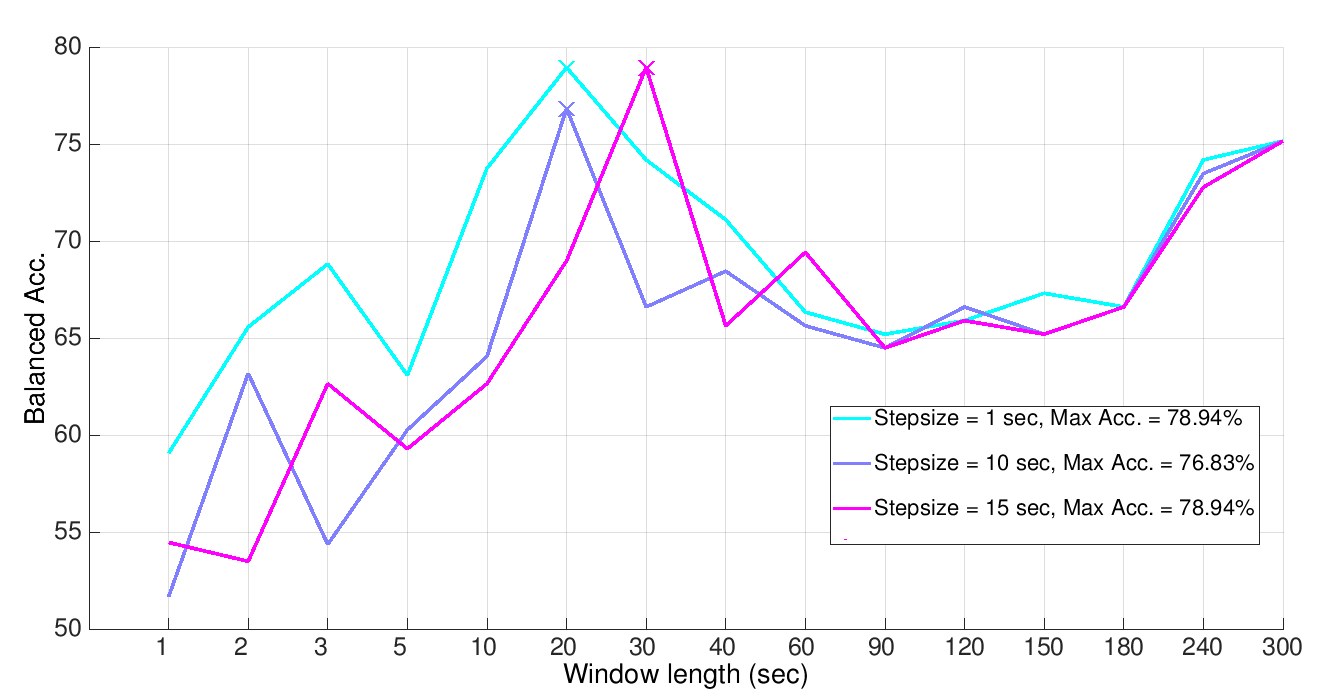}
    \caption[Online evaluation with different window and step size]{Online evaluation with different windows and step sizes. Balanced accuracy is calculated from LOO results, and the prediction of each session is made by a majority vote across all the sliding windows. The cyan, blue, and purple lines show the results of step sizes 1 sec, 10 sec, and 15 sec, respectively. The window length with maximum balanced accuracy is marked by a cross in all step sizes.}
    \label{online_vote}
\end{figure*}

We evaluate the online capability of our 11-channel LDA model with the IC projection artifact removal method.
We selected this model because of its potential to achieve comparable performance to the model processed by offline ICA with fewer recording channels.
Fig.~\ref{online_vote} shows the balanced accuracy of the 11-channel LDA model with different pairs of sliding window lengths and step sizes. 
When using a window length equal to 1 second, the balanced accuracy of the model hovers around 55\%. 
When the window length increases to 20 seconds, the balanced accuracy peaks at 78\%.
When the window length equals 5 minutes, the balanced accuracy drops to 75\%, and there is a 4\% decrease compared to offline results shown in Table~\ref{stress_lda_ch}.
Though the performance declined concomitantly with reductions in step sizes at window lengths of less than 20 seconds, we found that the difference in stress-detection performance with different step sizes was subtle, whereas the choice of window length dominated the model's performance.
Contrary to our expectations that a longer window length woudl provide better PSD estimation, the balanced accuracy actually declined when the window length is between 40 seconds and 3 minutes. 

\section{Discussion}
Mental stress can greatly affect one's cognitive performance.
Detecting stress in daily life can benefit clinical treatments and task performance, as mental stress can impact many aspects of a person's cognitive and physiological functions.
However, stepping into the real world outside of a research lab's controlled conditions brings several challenges. 
In this study, we addressed the following considerations that are key to successful real-world stress detection: (1) obtaining robust EEG signal through artifact removal, (2) online classification capabilities using the IC projection artifact removal method, and (3) optimizing the number of input EEG channels.
\par
Each step in our artifact removal pipeline incrementally improved the performance of our stress detection models, as shown in Table~\ref{stress_lda_artifact} and Table \ref{stress_nn_artifact}.
Moreover, our proposed IC projection artifact removal method yielded a balanced accuracy comparable to offline ICA.
These results suggest that there is limited room for improvement in further removing muscle artifacts after removing eye artifacts.
\par
Our proposed IC projection artifact removal method relies heavily on the consistency of eye-related ICs across recordings.
Fig.~\ref{nb_eyeIC_portion} shows that nearly 90\% of the sessions in this dataset contain 2 or more eye-related ICs, indicating that blink-related and saccade-related ICs appear in almost all sessions.

In addition to identifying blink-related and saccade-related ICs, we observed an extra IC in certain sessions that could potentially be linked to minor lateral eye movements or facial expressions such as frowning. Conversely, the ICs associated with muscle activity exhibited greater diversity. Due to the complex nature of muscles, comprising numerous fibers, capturing a single body movement within a single IC poses a challenge for ICA.


\par
When comparing the performance of the LDA and NN models, we consistently observe superior performance from the LDA model over the NN model. One plausible explanation for this difference is overfitting. The NN model offers many trainable parameters, yet the available training data is relatively limited. Upon comparing models' performances with and without SMOTE, we noted only marginal enhancements for both LDA and NN models when employing SMOTE. This could be attributed to the scarcity of training data, as SMOTE assumes that the sampled data shares similar statistical characteristics with the underlying population.


\par
Low-density channel arrays are preferred to reduce setup times and equipment costs. 
Since ICA can decompose eye activities with channels near the frontal region, a full scalp montage may be unnecessary, particularly when using the proposed IC projection artifact removal method.
Moreover, our LDA model achieved comparable accuracy using only 11 versus 30 channels of EEG data (Table \ref{stress_lda_ch}), in keeping with the idea that features from a limited subset of channels (Fz, FCz, and Cz) can capture the discriminative information for reliable stress detection.
However, because our IC projection artifact removal method is based on independent components, its performance drops with too few input channels—as shown by our finding that with only 5 channels, ICA fails to decompose eye activities into distinct blink- and saccade-related ICs.
Instead, ICs represent a mixture of brain and eye activities.
Assuming that the most likely blink- and saccade-related ICs would be removed from the analysis, only three ICs would remain to reconstruct the surface EEG signals.
Also, a substantial portion of brain signals would be lost with the "mixed" ICs accounting for eye activities, resulting in a decline in model performance.

\par
Finally, we evaluated the online capability of the 11-channel LDA model with the IC projection artifact removal method by investigating the effect of different pairs of sliding window lengths and step sizes on the balanced accuracy.
The choice of window length over step size dominates the model's performance, probably because of the quality of the PSD estimation.
The longer the window length, the better the estimation quality.
We observed the best performances across different step sizes using a window length of around 20 seconds, indicating that the quality of PSD estimation was sufficient to detect stress level changes.
When the window length was shorter than the step size, the algorithm ignored the data points between them, resulting in a down-sampling effect that hampered the performance of stress detection.
The effect is more severe when using a large step size, as observed in Fig.~\ref{online_vote}.
On the other hand, at a window length of 5 minutes, we found that the balanced accuracy dropped slightly compared to the offline results, possibly because some sessions involved an eyes-open resting baseline that exceeds 5 minutes. 
\par
One thing that caught our attention is that, instead of a monotonic increase, the model's balanced accuracy dropped before returning to expected performance levels when using a 5-minute window length.
We hypothesize that mental stress changes do not affect EEG signals constantly, but rather, on a repeated cycle such that EEG-based changes due to psychological stress might be diluted by the long window length.
However, obtaining the benchmarks for stress detection in an online scenario would make testing this hypothesis very difficult.

\section{Conclusion}
This study proposes a robust, online capable, EEG-Based mental stress detection algorithm suitable for pervasive, real-world monitoring using a low-density channel array.
To remove ocular artifacts in real time, we proposed an IC projection artifact removal method that reduced the required computational time while maintaining performance compared to offline ICA.
This study also explored the influence of different models, preprocessing pipelines, and the number of recording channels.
Our results show that with 15 features from Fz, FCz, and Cz channels, the proposed LDA-based stress detection algorithm can achieve a 78\% balanced accuracy in an imbalanced binary stress detection problem.
Moreover, for the purposes of artifact removal, the proposed LDA-based stress detection algorithm requires only 11 recording channels near the frontal region.
The findings of this study could lead to the deployment of daily mental stress monitoring in the future. 

\section{Acknowledgement}
Dr. Li-Wei Ko recorded the dataset used in this study at National Chiao Tung University (National Yang-Ming Chiao Tung University), Hsinchu, Taiwan.
The authors gratefully acknowledge the support from Dr. Li-Wei Ko, Microsoft Research, and the National Science Foundation.


\bibliographystyle{IEEEtran}
\bibliography{bib_EMBC2018}

\end{document}